\documentclass[manuscript]{acmart}

\graphicspath{{figures/}} 

\usepackage{subcaption}
\usepackage{multirow}
\usepackage{xcolor}

\usepackage{pifont}

\usepackage{mathtools}

\AtBeginDocument{%
  \providecommand\BibTeX{{%
    \normalfont B\kern-0.5em{\scshape i\kern-0.25em b}\kern-0.8em\TeX}}}

\copyrightyear{2024}
\acmYear{2024}
\setcopyright{rightsretained}
\acmConference[IUI Companion '24]{29th International Conference on Intelligent User Interfaces - Companion}{March 18--21, 2024}{Greenville, SC, USA}
\acmBooktitle{29th International Conference on Intelligent User Interfaces - Companion (IUI Companion '24), March 18--21, 2024, Greenville, SC, USA}
\acmDOI{10.1145/3640544.3645227}
\acmISBN{979-8-4007-0509-0/24/03}

\begin{document}

\title{Exploring AI-assisted Ideation and Prototyping for Choreography}

\author{Yimeng Liu}
\affiliation{
  \institution{University of California, Santa Barbara}
  \city{Santa Barbara}
  \postcode{93106}
  \country{USA}}
\email{yimengliu@cs.ucsb.edu}

\author{Misha Sra}
\affiliation{
  \institution{University of California, Santa Barbara}
  \city{Santa Barbara}
  \postcode{93106}
  \country{USA}}
\email{sra@cs.ucsb.edu}

\renewcommand{\shortauthors}{Liu and Sra}

\begin{abstract}
Choreography creation is a multimodal endeavor, demanding cognitive abilities to develop creative ideas and technical expertise to convert choreographic ideas into physical dance movements. Previous endeavors have sought to reduce the complexities in the choreography creation process in both dimensions. Among them, non-AI-based systems have focused on reinforcing cognitive activities by helping analyze and understand dance movements and augmenting physical capabilities by enhancing body expressivity. On the other hand, AI-based methods have helped the creation of novel choreographic materials with generative AI algorithms. The choreography creation process is constrained by time and requires a rich set of resources to stimulate novel ideas, but the need for iterative prototyping and reduced physical dependence have not been adequately addressed by prior research. Recognizing these challenges and the research gap, we present an innovative AI-based choreography-support system. Our goal is to facilitate rapid ideation by utilizing a generative AI model that can produce diverse and novel dance sequences. The system is designed to support iterative digital dance prototyping through an interactive web-based user interface that enables the editing and modification of generated motion. We evaluated our system by inviting six choreographers to analyze its limitations and benefits and present the evaluation results along with potential directions for future work. 
\end{abstract}

\ccsdesc[500]{Human-centered computing~Interactive systems and tools}

\keywords{AI-supported choreography, human-AI interaction}

\maketitle

\section{Introduction}
Choreography creation is a complicated process that demands significant cognitive and physical capabilities~\cite{stevens2003choreographic, kogan2002careers}. Choreographers develop ideas based on their knowledge base, experience, artistic taste, and choreographic themes to craft appealing choreography~\cite{ciolfi2016choreographers}. Additionally, they utilize their technical expertise to translate these choreographic ideas into physical dance movements. A harmonious fusion of creativity and physical execution is essential in creating dance pieces that effectively convey choreographers' creative vision and engage audiences~\cite{kirsh2009choreographic, fdili2017seeing}. 

To assist choreography creation, prior research has designed non-AI systems and AI approaches to support the cognitive and physical requirements. 
Non-AI systems have been proven to be effective in analyzing dance movements for a deeper understanding and improved execution of the movements~\cite{oulasvirta2013information, velloso2013motionma}. These systems have also helped to extend dancers' physical expressivity to inspire novel concepts and motivate new forms of dance choreography and performance~\cite{eriksson2019dancing, ciolfi2018knotation}. On the other hand, AI choreography-support methods have allowed the generation of dance movements and sequences for ideation based on various input modalities, including text~\cite{gong2023tm2d}, music~\cite{chuang2022music2dance}, and video~\cite{chan2019everybody}. 

According to a prior study~\cite{calvert1989composition}, motivating novel concepts, reducing physical demands, and allowing iterative prototyping all contribute to the success of choreography creation. This is mainly due to the time constraints applied to the choreography creation process, requiring wise management of effort and time and ensuring productivity within the limited timeframe. Although existing non-AI systems have shown effectiveness for dance movement analysis and inspiration, they fall short of saving physical effort due to heavy reliance on human prototyping to craft and adjust dance movements. Additionally, prior AI methods have been leveraged to support ideation through AI-generated content; however, most of them fail to support iterative prototyping of the outcomes -- a crucial demand to polish choreographic ideas and dance movements~\cite{ciolfi2016choreographers}. 

To fill this research gap, we introduce an AI choreography-support system that integrates a generative AI model to facilitate ideation and digital prototyping of choreography materials. First, the system allows the generation of diverse dance sequences based on textual descriptions for ideation. Second, we address the limitations of prior AI algorithms by supporting the iterative editing of AI-generated outcomes through a web-based user interface. By interacting with the interface, users can digitally prototype choreographic ideas and materials without requiring physical prototyping. Moreover, the system enables the documentation of intermediate and refined results, capturing users' attempts and final products in videos and 3D animated motions. In conclusion, we present the opportunities and limitations identified through a system evaluation and outline future directions based on the insights obtained from the evaluation results.

\section{Related Work} \label{sec:related_work}
\subsection{Non-AI Choreography-support Systems}
In a recent review paper~\cite{zhou2021dance}, existing non-AI choreography-support systems have been categorized based on their intent to enhance human body expressivity and facilitate the analysis of body movements. 
First, choreographers have utilized computing technologies to augment the human body expressive capabilities~\cite{karpashevich2018reinterpreting, johnston2015conceptualising, janauskaite2019establishing, eriksson2019dancing, raheb2018choreomorphy, jochum2019tonight, gemeinboeck2017movement, ladenheim2020live}. 
Further, choreographers have embraced motion capture and 3D modeling techniques to analyze and understand abstract physical movement characteristics in dance~\cite{oulasvirta2013information, velloso2013motionma, singh2011choreographer, alaoui2012movement, akerly2015embodied, carlson2015moment, carlson2015sketching, ciolfi2018knotation}. 
Additionally, technologies have been specifically developed to support creative processes in choreography, such as ideation~\cite{molina2017delay, raheb2018choreomorphy, eriksson2019dancing} and documentation~\cite{ciolfi2018knotation, zhou2023here}.
While these systems can effectively contribute to the analysis of dance movements, the expressive enhancement of dancers, and the stimulation of novel ideas, they depend on the active physical involvement of choreographers and dancers for choreography creation, refinement, and performance. This demand can be challenging, particularly due to time constraints, as choreographers must wisely distribute their effort and time across various tasks during the choreography creation process, such as thinking, communicating, and physical prototyping, to achieve success while minimizing stress. 

\subsection{AI Choreography-support Methods}
AI choreography-support methods have proven useful in producing choreography materials for ideation during choreography creation. 
These methods utilize generative AI algorithms, such as generative adversarial networks and diffusion models, to generate dance movements and sequences based on various input modalities. One of the input options that have been extensively explored is music-conditioned dance generation~\cite{chuang2022music2dance, zhang2022music, alexanderson2023listen}. However, the initiation of choreography creation does not necessarily require music~\cite{mason2012music}; choreographers may even intentionally refrain from using music to prevent excessive reliance on it during dance creation~\cite{stevens2009moving}.
Alternatively, text has been used as a useful means to convey choreographic ideas for dance making~\cite{gong2023tm2d}. Moreover, dance videos encode visual information acted by dancers and have been utilized to drive the generation of new dance sequences~\cite{chan2019everybody}. 
These input options can effectively generate dance and contribute to a rich set of materials for choreographers to incorporate into their dance pieces. These materials also have the potential to stimulate creative thinking by encouraging the recombination and exploration of materials in innovative ways. 
Despite the advantages of existing AI choreography-support methods, they usually fail to allow users to edit outcomes, limiting them to one-time automatic dance generation. We address this limitation by enabling the editing of AI-generated results to align with the iterative choreography process.

\section{System} 

\begin{figure*}[!ht]
    \centering
    \includegraphics[width=\textwidth]{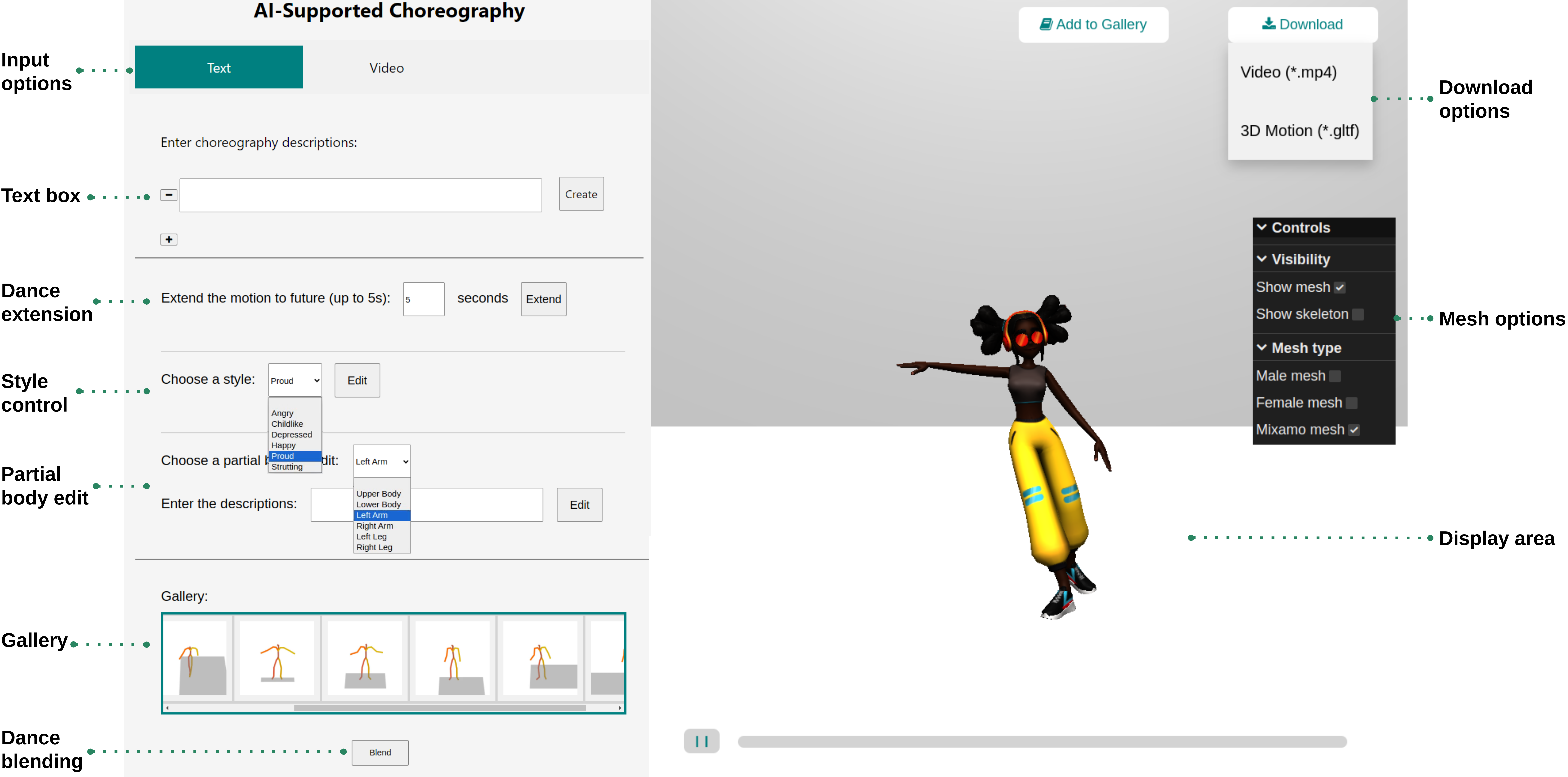}
    \caption{User interface. The interface accepts text and video as input. Typing text descriptions leads to dance generation, and uploading videos offers converted 3D dance sequences for editing. After dance sequences are generated or uploaded, users can further edit them using the editing options. Dance extension is allowed by typing in the length to extend by up to 5 seconds. For style control, users can pick styles from a drop-down menu. To edit partial body movements, users can choose a body part and describe how they want to change it in the text box below. After creating dance sequences, users can add them to the \textit{Gallery} for future use, shown as thumbnails. If users select dance sequences from the \textit{Gallery}, they can blend them. On the right, users can view and interact with the generated dance sequences represented by a digital avatar. The visibility of the avatar's mesh and skeleton is adjustable via the checkboxes. Three types of avatar meshes are available in our current prototype: SMPL~\cite{SMPL_2015} male, female, and Mixamo~\cite{mixamo} mesh. Lastly, users can download generated dance sequences as 2D videos and 3D animated meshes.}
    \Description{The user interface is divided into user interaction options on the left and a result display area on the right. From top to bottom, the interaction options include input options, a text box if the input option is text, and a video upload button if the input option is video. Below are the edit options: a dance extension length input box, a style selection dropdown menu, a partial body edit dropdown menu together with a description text box, a gallery containing user-created dance sequences, and a blend button. In the display area, generated dance sequences performed by a virtual avatar are shown in the center. The top right features buttons for adding to the gallery and downloading, while the right side includes mesh visibility and choice options. The bottom displays a progress bar with a pause/play button.}
    \label{fig:user_interface}
\end{figure*}

\subsection{User Interface} \label{sec:user_interface}
Figure~\ref{fig:user_interface} shows the user interface. Features on the user interface are motivated by a formative study involving seven choreographers. 
The left side presents two tabs for user input: text and video. Opting for the text tab provides a text box for users to describe dance movements, with a + button allowing the addition of more input boxes as needed. Below this, an editing panel enables users to modify generated content: expand dance sequences, alter dance styles, change partial body movements, and blend two dance sequences. Choosing the video tab replaces the text input box with a file upload button to import dance videos while the editing panel remains unchanged. By combining textual and visual content, our system maximizes the potential to convey creative ideas, particularly when using a single mode of expression is insufficient. 
The right side is the display area. To visualize dance sequences, users can switch the mesh type to a SMPL~\cite{SMPL_2015} male, female, or Mixamo~\cite{mixamo} mesh through the corresponding checkboxes. The meshes are interactive, allowing rotation, scaling, and movement. The visibility of the meshes and skeleton is also adjustable using checkboxes. The \textit{Add to Gallery} button saves user creations into the \textit{Gallery} for future reference. The \textit{Download} button permits exporting the animated mesh in \textit{.gltf} format for use in 3D scenarios and in \textit{.mp4} format as a video. 

\subsection{Functionality}
\paragraph{\textit{\textbf{Dance Generation}}} \label{sec:dance_generation}
Users can input a dance genre or specify dance movements as foundational components to obtain three dance sequences each time. 
Generating diverse dance sequences based on a text description leverages the probabilistic nature of the generative AI model we employ. 
The system generates dance sequences up to 10 seconds. On average, generating three outputs takes 10 seconds, with the time varying based on internet latency. 

\textit{\textbf{Dance Extension}}. \label{sec:dance_extension}
Dance extension empowers users to expand sequences they upload from videos or those the system generates. This feature facilitates improvisation from existing data, aligning with a prevalent choreography method favored by many choreographers~\cite{ciolfi2016choreographers}. It also allows our system to overcome duration constraints imposed by the length of motion sequences used to train the underlying AI model. To surpass the 10-second limit set by this model, we permit a 5-second extension for each user-initiated extension. This decision emerged from experimentation, revealing that a 5-second extension maintains the quality of the extended portion. Longer extensions led to motionless segments, while shorter ones failed to capture the entirety of the dance movements. Users can repetitively extend sequences through this process to obtain longer durations. 

\textit{\textbf{Style Control}}. \label{sec:style_control}
Our system facilitates the editing of dance sequence styles, offering options including angry, childlike, depressed, happy, proud, and strutting. Incorporating these six styles serves as a proof-of-concept pipeline to introduce emotional changes in dance movements.
These styles were initially introduced by SinMDM~\cite{raab2023single} as part of a motion harmonization special case. Given that the HumanML3D dataset~\cite{guo2022generating} used to train the underlying AI model lacks style labels, directly describing emotions in text prompts does not inherently modify dance styles. To address this, we explicitly defined the six styles based on the HumanML3D dataset and trained models for each type of emotion-to-motion style transfer using a similar training method as SinMDM~\cite{raab2023single}. 

\textit{\textbf{Partial Body Edit}}. \label{sec:partial_body_edit}
Minor adjustments, such as modifications to partial body motions, are often necessary in the choreographic process. The system allows changing the movements of the upper and lower body, arms, and legs.
As an editing feature, changes made to partial body movements allow users to observe how the remaining body movements adjust according to the edited segment. Alternatively, partial body edits can serve as a dance generation feature. For instance, using a prompt like ``A person is doing a moonwalk; restrict movements of arms'' generates a dance sequence adhering to the specified constraint on arms. This shows how partial body control directly influences the dance generation process, presenting a distinction from the effects observed during the editing phase. 

\textit{\textbf{Dance Blending}}. \label{sec:dance_blend}
Users can concatenate two dance sequences to obtain a longer one that seamlessly merges the chosen sequences, incorporating a 5-second connecting segment to ensure a smooth transition. 
Blending dance sequences empowers users to expand creative outcomes and explore novel combinations of dance genres, e.g., deviating from classic dance styles by blending multiple dance genres.
This becomes particularly relevant as mastering multiple dance genres and mixing them for choreography creation is a non-trivial task for humans. 

\textit{\textbf{Documentation}}. \label{sec:documentation}
Users can export 2D documentation containing generated dance sequences before and after edits in a \textit{.mp4} format. These videos showcase animated human skeletons executing the generated dance movements. Additionally, users can download the generated dance movements in 3D, stored in a \textit{.gltf} format. This 3D representation can be used in various 3D scenarios, such as AR/VR/MR. 
Supporting both 2D and 3D documentation offers clear advantages. Videos are easy to store and share; however, they only display one perspective at a time, making it challenging to observe hidden body movements. 3D animations address this issue by enabling users to view dance movements from different angles. Moreover, our system automatically saves text prompts used to generate and edit dance sequences, allowing users to keep track of their attempts within the system. 

\subsection{Example Results}
\begin{figure*}[!ht]
    \centering
    \begin{subfigure}[b]{0.75\textwidth}
        \includegraphics[width=\textwidth]{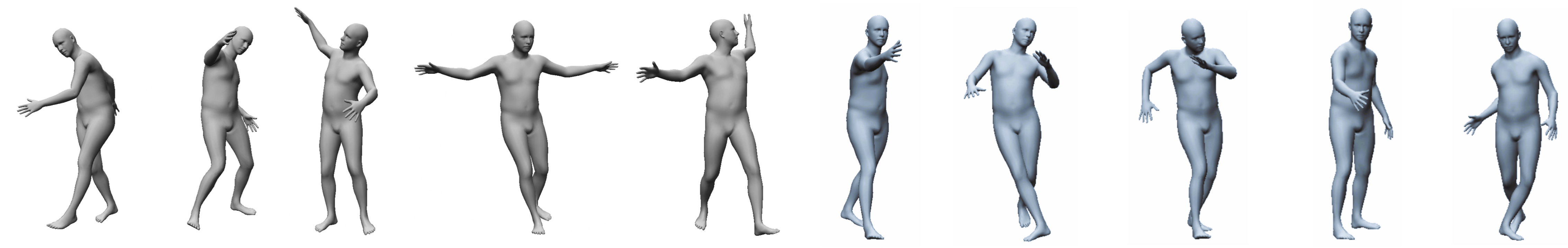}
        \caption{Dance extension of ``A man is dancing ballet''.}
        \label{fig:extended_sequence}
    \end{subfigure}
    \begin{subfigure}[b]{0.35\textwidth}
        \includegraphics[width=\textwidth]{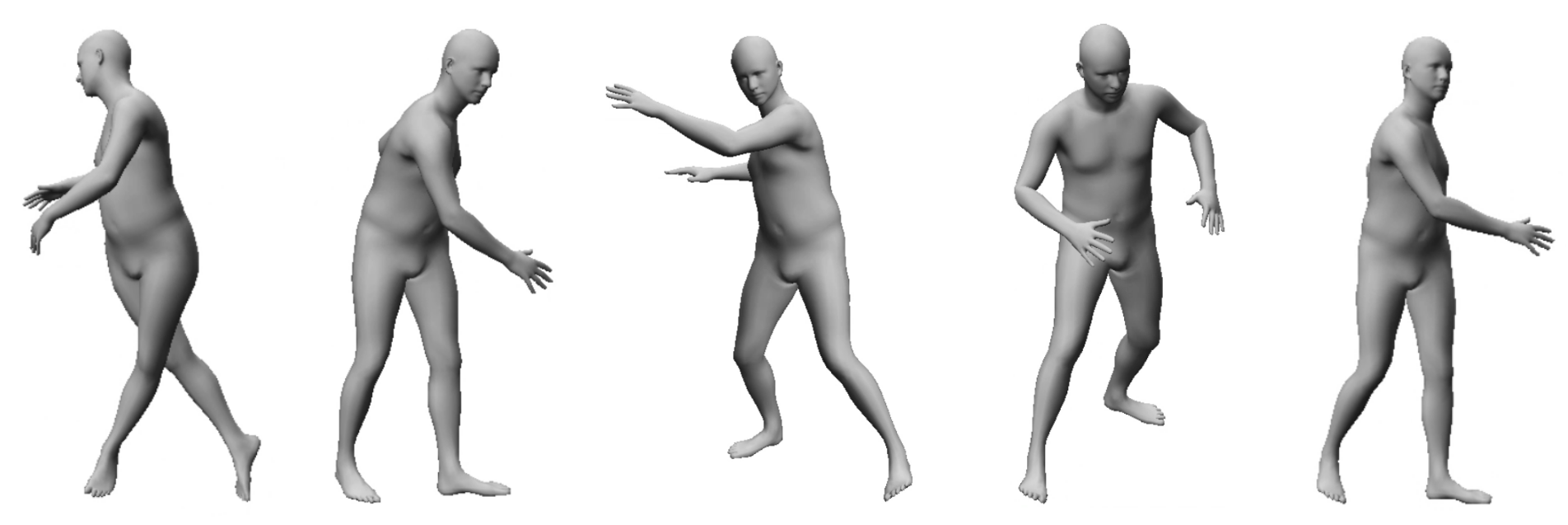}
        \caption{Style control: angry.}
        \label{fig:style_control_angry}
    \end{subfigure}
    \hspace{0.05\textwidth}
    \begin{subfigure}[b]{0.32\textwidth}
        \includegraphics[width=\textwidth]{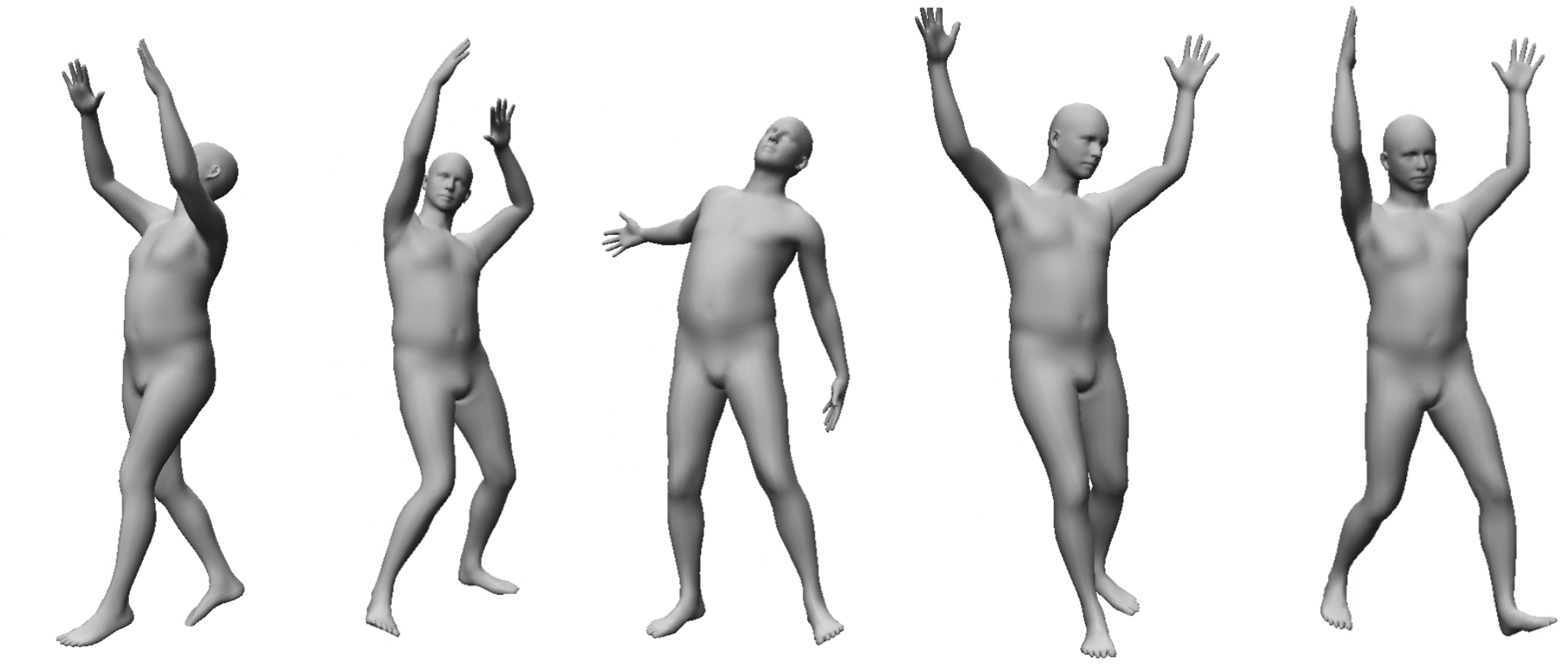}
        \caption{Partial body: ``Keep the arms raised up''.}
        \label{fig:upperbody_armsup}
    \end{subfigure}
    \begin{subfigure}[b]{0.34\textwidth}
        \includegraphics[width=\textwidth]{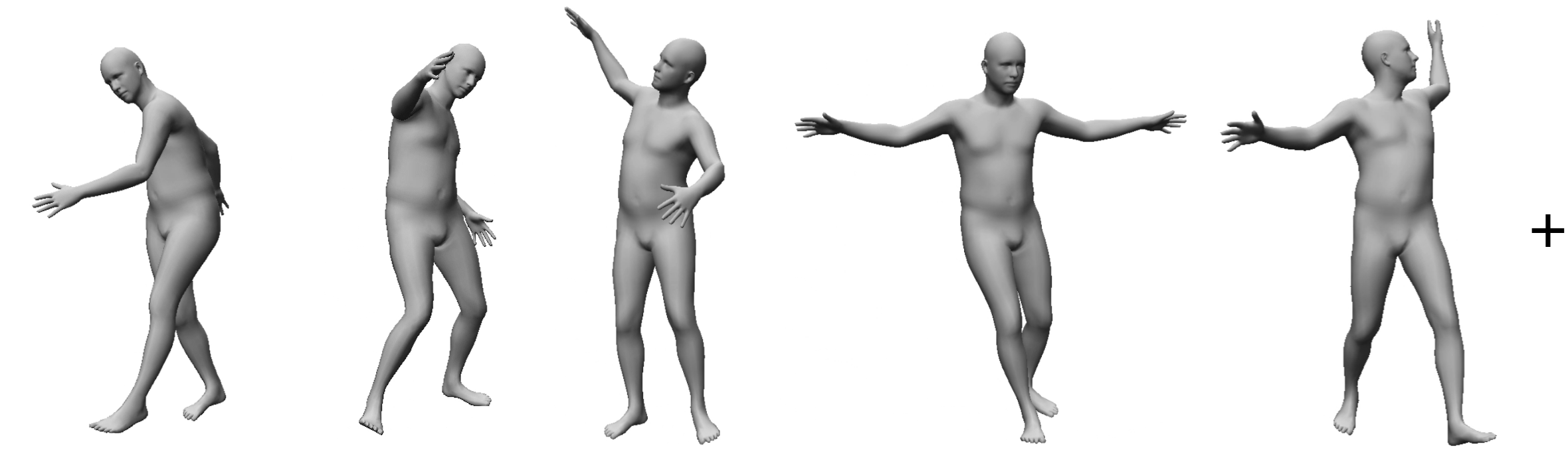}
        \caption{``A man is dancing ballet''.}
        \label{fig:blend_original_sequence}
    \end{subfigure}
    \begin{subfigure}[b]{0.3\textwidth}
        \includegraphics[width=\textwidth]{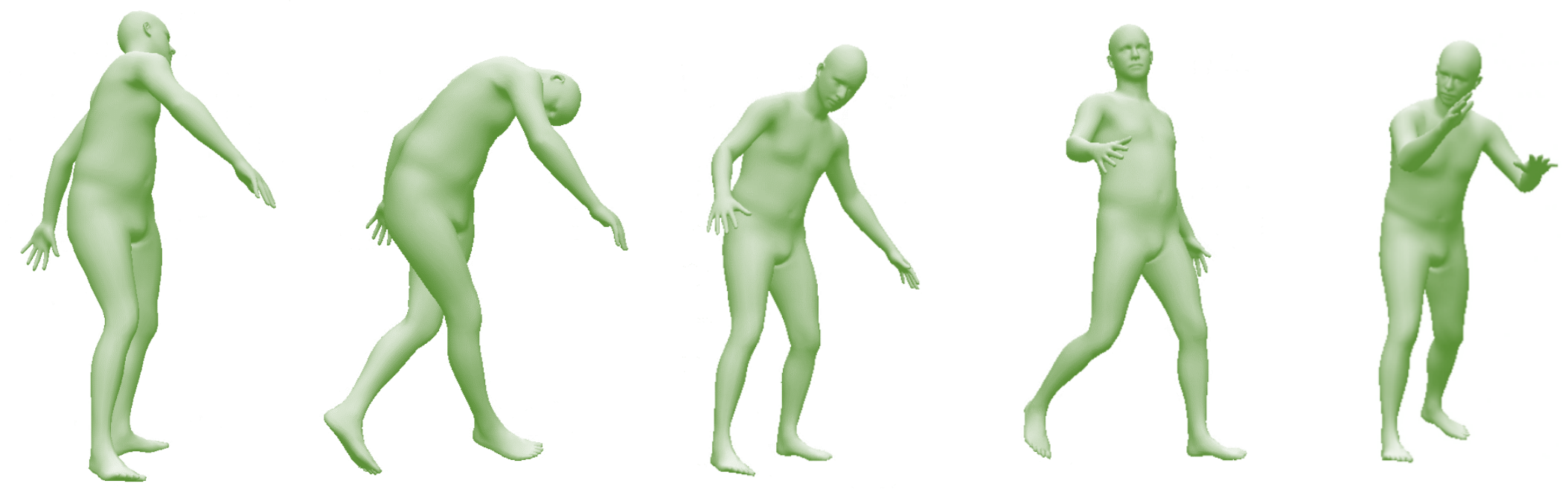}
        \caption{Connecting dance sequence.}
        \label{fig:blend_connect_sequence}
    \end{subfigure}
    \begin{subfigure}[b]{0.31\textwidth}
        \includegraphics[width=\textwidth]{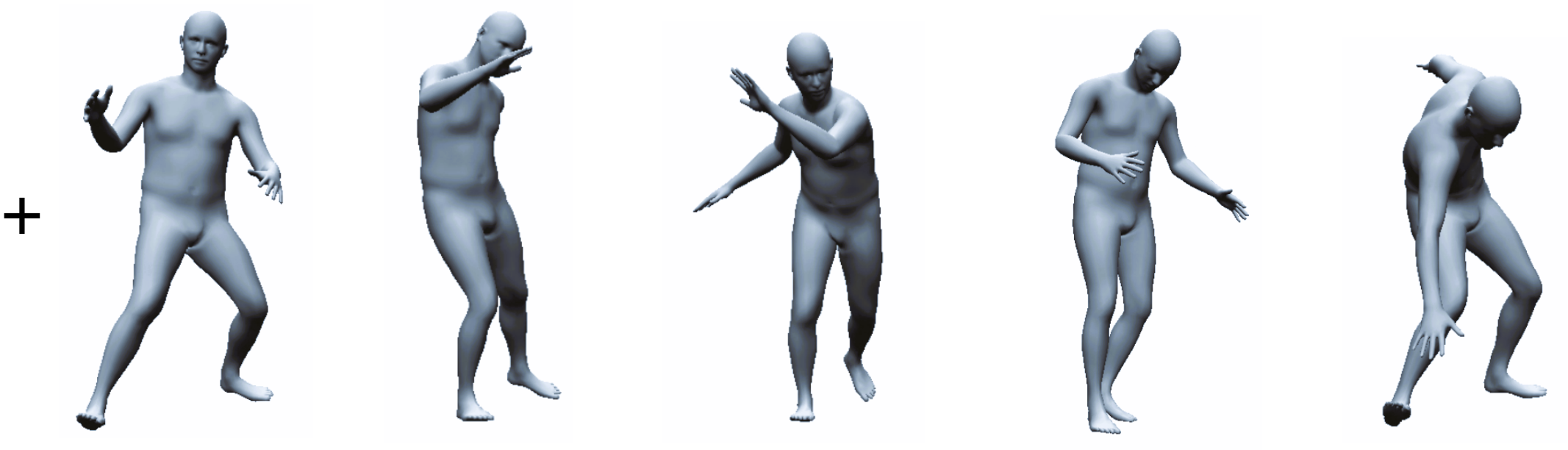}
        \caption{``A man is dancing hip-hop''.}
        \label{fig:blend_second_sequence}
    \end{subfigure}
    \caption{Results of a dance sequence generated from ``A man is dancing ballet'' and edited sequences based on it. (a) The original dance sequence is 10 seconds, shown in gray, and the extended segment by 5 seconds is blue. (b) The style ``angry'' is applied to the original dance sequence. (c) The partial body movements are altered according to ``Keep the arms raised up''. (d) - (f) The blending results of the original sequence and a 10-second dance sequence generated from ``A man is dancing hip-hop'' in blue, with a 5-second connecting dance sequence merging them in green.}
    \Description{The dance editing results are shown as five frames for each dance sequence using the SMPL male mesh.}
    \label{fig:editing_results}
\end{figure*}

Figure~\ref{fig:editing_results} presents a few results obtained using our system, including a 10-second dance sequence generated from ``A man is dancing ballet'' and edited results based on this sequence using the editing options. Each dance sequence is shown as five frames executed by the SMPL male mesh~\cite{SMPL_2015}. 

\subsection{Technical Details} \label{sec:technical_details}

\begin{figure*}[!ht]
    \centering
    \includegraphics[width=0.75\textwidth]{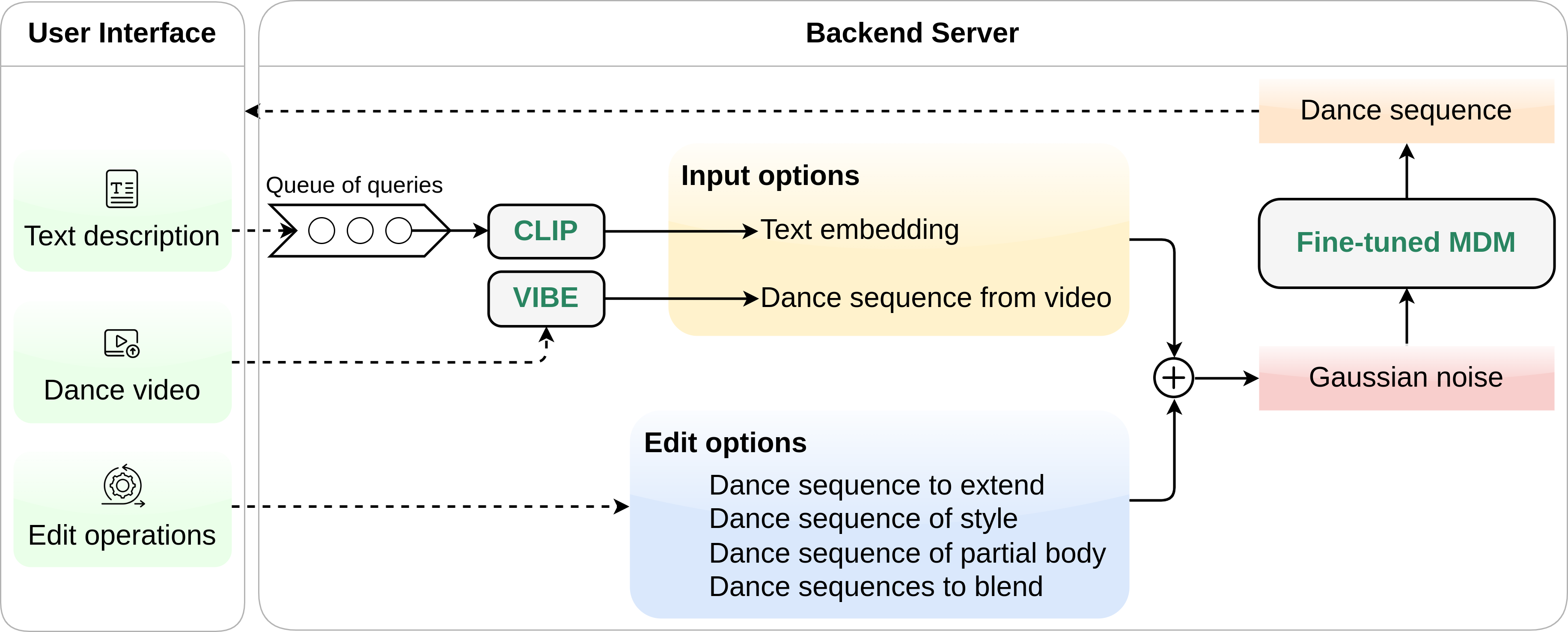}
    \caption{Technical details. The system's user interface and backend server are connected via a TCP socket, shown as dashed arrows. The user interface sends users' text descriptions, uploaded videos, and editing operations to the backend and receives generated dance sequences for display and interaction. The backend server encodes each text description from the query queue with CLIP~\cite{radford2021learning} embeddings and converts dance movements from the video files into 3D sequences using VIBE~\cite{kocabas2020vibe}. These inputs and user-initiated edit operations are fed into the fine-tuned MDM~\cite{tevet2023human} to generate dance sequences. Dataflows within the backend server are represented as solid arrows.}
    \Description{This figure consists of the user interface and the backend server. Within the user interface, three types of user interaction are depicted from top to bottom: text description, dance video, and edit operations. In the backend server, the input options include encoded CLIP text embeddings and 3D dance sequences converted from videos, while the editing options include the edit operations. The input and edit options are summed with a Gaussian noise sequence before input into the fine-tuned MDM to produce a sequence of dance movements.}
    \label{fig:technical_details}
\end{figure*}

\paragraph{\textit{\textbf{Frontend User Interface and Backend Server}}}
The frontend user interface is constructed using HTML/CSS and JavaScript. It communicates with the backend server through a TCP socket~\cite{tcp_server}. Figure~\ref{fig:technical_details} shows the frontend and backend connections via dashed arrows and the operations within each of them. 

When users choose text as input, it is encoded as a query, placed in a queue, and transmitted to the backend. The backend processes this queue -- extracts all the queries, encodes them into text embeddings using CLIP~\cite{radford2021learning}, and passes the embeddings to the generative model for dance sequence generation. 
Alternatively, the video file is sent to the backend if users upload a dance video as input. The video is processed by VIBE~\cite{kocabas2020vibe} to track the dancing person and obtain a 3D dance sequence. 
Each dance sequence is identified with a unique ID; the IDs are updated if any dance sequence is edited. Specifically, if a dance sequence is picked to edit, the frontend sends the editing operations, and the backend loads the dance sequence to be edited through its ID and passes it into the generative model.
The resulting dance sequences, generated or edited by users, are returned to the user interface for display and interaction. We use the three.js library~\cite{threejs} to load 3D meshes and animation parameters into a web browser. We employ SMPL meshes~\cite{SMPL_2015} in both female and male types, along with a Mixamo mesh~\cite{mixamo}, as avatars to showcase the generated dance sequences.

\textit{\textbf{Generative AI Model}}.
We use the Motion Diffusion Model (MDM)~\cite{tevet2023human} as the underlying AI model for dance generation and editing. This model was initially pre-trained on the HumanML3D dataset~\cite{guo2022generating}, covering a large number of natural language descriptions for various daily human motions. 
We further fine-tuned this model using the AIST++ dance dataset~\cite{Li_2021_ICCV}. In this process, the authors manually labeled the AIST++ dataset with the ten available dance genres, providing text descriptions for fine-tuning. The fine-tuning took approximately 13 hours on a machine equipped with a single NVIDIA RTX 3090 graphics card. The fine-tuned model allows dance sequence generation and editing in the model inference phase. In the backend server module shown in Figure~\ref{fig:technical_details}, we illustrate the input and edit options and the inference process to generate dance sequences using the fine-tuned model.

\section{System Evaluation}
To help us understand our system's ability to support professional choreographers, we invited six participants (five females and one male), each assigned an ID from P01 to P06, for 1:1 system evaluations. The participants had four to over ten years of choreography experience in hip-hop, jazz, contemporary, and ballet for dance instruction and commercial performance. The study was conducted following approval from the local Institutional Review Board (IRB) under protocol \# 21-23-0275. Participants provided us with informed consent and underwent participant training before the evaluation. During the study, they used the system for 20 minutes to create multiple dance sequences as they wanted. Following their interaction with the system, we conducted semi-structured interviews with each participant to understand their experience and gather feedback. After the study, we performed a reflexive thematic analysis~\cite{braun2006using, braun2012thematic} on the interview data. This section outlines the insights extracted from our data analysis. 

\subsection{AI-assisted Choreography Ideation}
Participants found the varied outcomes produced by the system offered them choreographic materials and could potentially spark their new thinking for dance making. For instance, they commented that the AI-generated dance sequences allowed them to explore ``\textit{a broad scope of materials}'' (P01) that they can adapt ``\textit{for future use or inspire me to think of new moves or combinations}'' (P01). Moreover, they thought that ``\textit{the diverse results could help me be more efficient as I could simply recombine them to get a new dance sequence}'' (P02). 

Some participants raised concerns regarding the efficiency of using a generative AI system vs. a search engine to collect materials for choreography ideation, with comments like ``\textit{compared with searching for online videos, one limitation of this system is searching online is possibly more efficient than using AI to generate results. The former takes a few seconds to obtain hundreds of results, but the latter offers only three outputs}'' (P06). This feedback raises a question of how in-demand vs. a high volume of results affects choreography ideation. Both methods could be time-consuming to obtain useful materials, with the former requiring prompt engineering and the latter demanding careful selection. A harmonious combination of the two approaches may enhance the development of future ideation techniques, e.g., using AI-generated content for more precise searches or incorporating searched results for AI-driven polishment. 

\subsection{AI-assisted Choreography Prototyping}
Participants provided feedback on the efficacy of using the editing options to modify and refine dance movements. For example, they found that ``\textit{the editing options are helpful to speed up early prototyping of ideas using my computer}'' (P04), and the documented dance sequences allowed them to ``\textit{save results for myself to revisit}'' (P05) and ``\textit{update based on what I left from previous tries}'' (P04). 

Although digital prototyping assisted by an AI system could save time and physical effort for early testing of choreographic ideas, some participants commented that the digital format ``\textit{could only reveal dance movements but failed to convey emotion and energy}'' (P03) through the movements. Implicit information, including emotion, is crucial for dance performance and creation~\cite{kirsh2009choreographic}. While our system utilizes emotion as a style condition for choreography modification, this feedback suggests a need for improved digital tools to convey information implicitly incorporated in choreography through facial expressions~\cite{audio2face} or muscle motions~\cite{zheng2021deep} of a digital avatar for more accurate and thorough visualization and prototyping of both implicit and explicit choreographic ideas.

\section{Conclusion and Future Work}
This paper introduces an interactive AI-based system designed to support choreography ideation to stimulate new concepts and digital prototyping to reduce physical demands for choreographic idea testing and polishment. Our system features an interactive web interface and a generative AI-powered backend, enabling fast generation, iterative editing, and 2D and 3D documentation of dance sequences. We obtained initial feedback on the system's strengths and weaknesses from a system evaluation. We aim to refine the system in our future work and gain deeper insights through extensive evaluations and discussions to inform the development of future AI-based choreography-support systems.

\begin{acks}
We want to thank Jennifer Jacobs for offering us valuable insights regarding the system evaluation design and members of the Human-AI Integration Lab and Expressive Computation Lab at UCSB for their feedback on the system functionality. 
\end{acks}

\bibliographystyle{ACM-Reference-Format}
\bibliography{iui2024}

\end{document}